\documentclass[twocolumn,showpacs,prl,10pt]{revtex4}
\usepackage[english]{babel}
\usepackage{amsmath,amssymb} 
\usepackage{times}
\usepackage{epsfig} 
\begin{document} 
\title{Incoherent transport induced by a single static impurity in 
a Heisenberg chain}

\author{O. S. Bari\v si\' c$^{1,2}$, P. Prelov\v sek$^{1,3}$}
\affiliation{$^1$J.\ Stefan Institute, SI-1000 Ljubljana, Slovenia}
\affiliation{$^2$Institute of Physics, HR-10000 Zagreb, Croatia}
\affiliation{$^3$ Faculty of Mathematics and Physics, University of
  Ljubljana, SI-1000 Ljubljana, Slovenia}
\author{A. Metavitsiadis$^4$, X. Zotos$^4$,} 
\affiliation{$^4$
  Department of Physics, University of Crete and Foundation for
  Research and Technology-Hellas, P.O. Box 2208, 71003 Heraklion,
  Greece}

\date{\today}
                   
\begin{abstract}
The effect of a single static impurity on the many-body states and on
the spin and thermal transport in the one-dimensional anisotropic
Heisenberg chain at finite temperatures is studied.  Whereas the pure
Heisenberg model reveals Poisson level statistics and dissipationless
transport due to integrability, we show using the numerical approach
that a single impurity induces Wigner-Dyson level statistics and at
high enough temperature incoherent transport within the chain, whereby
the relaxation time and d.c. conductivity scale linearly with
length.
 
\end{abstract}

\maketitle 

Transport in the one-dimensional (1D) quantum system of interacting
particles still offers several fundamental theoretical challenges. In
the context of the electron transport through barriers or weak links
the role of repulsive electron-electron interactions within the wire
was shown to be crucial \cite{kane,furu} since at low temperature $T
\to 0$ the interaction within Luttinger-liquid (LL) phenomenology
renormalizes the transmission through the barrier to zero effectively
cutting the chain for transport \cite{mede}. Neglecting the effect of
Umklapp processes within the wire above the Kane-Fisher temperature
$T^*$ the transmission through the barrier should become finite
\cite{kane,furu} indirectly confirmed in a numerical study of the 1D
spin model \cite{romm}.  On the other hand, in the last decade it has
become increasingly evident that the LL low-energy description is not
enough to establish transport properties, if they are dominated by
Umklapp processes. It has been shown \cite{cast,znp} that pure 1D
integrable models of interacting fermions exhibit in spite of Umklapp
at any $T>0$ dissipationless (ballistic) transport manifested, e.g.,
in a finite charge stiffness $D(T>0)>0$. On contrary, a generic system
of interacting fermions would exhibit dissipation in the wire and also
finite d.c.  transport coefficients, e.g., the d.c. conductivity
$\sigma(\omega \to 0)=\sigma_0 <
\infty$. The distinction is closely linked to the statistics of
many-body levels \cite{brod}, which follow the Poisson level
distribution for the integrable system and the Wigner-Dyson (WD)
distribution for the generic nonintegrable system \cite{cast}. Random
disorder, strong enough to overcome finite-size effects, in such
models generally leads to the WD statistics for nearest levels
\cite{jacq,rabs}, while the d.c. transport seems to be
normal (dissipative) at $T>0$ \cite{kara}, in contrast to the
Anderson-type localization persisting at $T=0$ \cite{schm}.

Our goal is to understand within this context the effect of a single
static impurity in the 1D integrable system of interacting
particles. While of fundamental importance this question is also
directly relevant in connection with ongoing experiments on novel
quasi-1D materials where electronic properties of the pure system can
be well described within the integrable spin-$1/2$ Heisenberg model
\cite{hess} with dilute impurities introduced in a controlled manner and
their influence studied, e.g., on the thermal transport.

In the following we show on the example of the 1D spin model with
periodic boundary conditions that a single static impurity can
qualitatively change the level statistics to the WD one, noticed also
in the recent study of the onset of quantum chaos \cite{sant} although
the perturbation scales as $1/L$ where $L$ is the length of the
system. At the same time, the impurity leads to the vanishing of the
spin stiffness $D$ at elevated $T>0$. In such a situation, it is
meaningful to discuss the decay of the spin and energy current within
the ring and related transport rates $1/\tau$ which we show to be well
defined and scale as $1/\tau \propto 1/L$ as expected for the
homogeneous wire with a single localized perturbed region.

We consider the 1D anisotropic Heisenberg model (AHM) with a
single-site static impurity field,
\begin{equation}
H=\sum_l J (S^x_{l+1}S^x_l+S^y_{l+1}S^y_l+\Delta S^z_{l+1}S^z_l) +
b_0 S^z_0, \label{ahm} 
\end{equation}
where $S^{\alpha}, \alpha=x,y,z$ are spin-1/2 operators, $J$ is the
magnetic exchange coupling (we use units $J=1$), $\Delta$ the
anisotropy and $b_0$ the local impurity field. Numerically we study
chain (ring) of length $L$ with periodic boundary conditions.

In the absence of the impurity the AHM, Eq.~(\ref{ahm}), is integrable
and as the consequence reveals the Poisson level distribution $P_P(s)=
\exp(-s)$ where $s=(E_{n+1}-E_n)/\Delta_0$ ($\Delta_0$ is the average
level spacing) as well as the dissipationless transport
\cite{znp}. Let us first consider the effect of finite $b_0$ on the
level statistics. We investigate this question by performing the
(full) exact-diagonalization (ED) study of finite size systems with
$L=10 - 16$. It should be pointed that in this range the number of
many-body states varies (in the $S^z_{tot}=0$ sector) in a wide range
$N_{st} = 10^2 - 10^4$ and the corresponding
$\Delta_0=2.10^{-2}-5.10^{-4}$. The general conclusion is that finite
$b_0>0$ induces WD distribution $P_{WD}(s)=(\pi s/2)\exp(-\pi s^2/4)$
following the random-matrix theory (RMT) \cite{brod} in spite of the
fact that the perturbation is only an $1/L$ effect (the perturbation
$b_0/2$ relative to the full energy span $\Delta_E \sim L J$).

To be concrete we present here two standard tests for the closeness of
the RMT.  The first one is parameter $\eta$ \cite{jacq} measuring
the normalized distance to the WD distribution,
\begin{equation}
\eta=\int_0^{s_0}[P(s)-P_{WD}(s)]ds /
\int_0^{s_0}[P_P(s)-P_{WD}(s)]ds, \label{eta}
\end{equation}
where $P(s)$ is the actual level distribution and $s_0=0.473$ is
chosen to be the intersection of $P_P(s)$ and $P_{WD}(s)$ \cite{jacq}.
In order to stay within the regime of homogeneous density of states we
analyze only one half of intermediate many-body states, as relevant
for the high-$T$ properties discussed here. ED results for resulting
$\eta$ as a function of $b_0$ for chosen intermediate $\Delta=0.8$ are
presented for different $L$. To avoid the effect of higher degeneracy
of levels at $S^z_{tot}=0, b_0=0$ presented results in Fig.~1 are for
$S^z_{tot}=1$. In the absence of impurity ($b_0=0$) we obtain $\eta=1$
since $P(s)=P_P(s)$ due to the integrability of the pure AHM. The most
important conclusion is that rather weak impurity $b_0 \sim 0.2$ in
largest $L=16$ causes a fast drop to $\eta \sim 0$, i.e., to $P(s)
\sim P_{WD}(s)$, whereby the threshold value of $b_0$ is decreasing
with $L$ so that for largest $L=16$ reachable with ED we get $P(s)
\sim P_{WD}(s)$ in the range $0.2<b_0<1.5$. On the other side, it is
quite remarkable that $\eta$ starts to recover towards $\eta \sim 1$
again for large $b_0\gg 1$. This can be easily explained by noting
that large $|b_0| \gg 1$ effectively cut the ring and lead to the AHM
with open ends which is again an integrable model.

\begin{figure}[htb]
\includegraphics[angle=0, width=.75\linewidth]{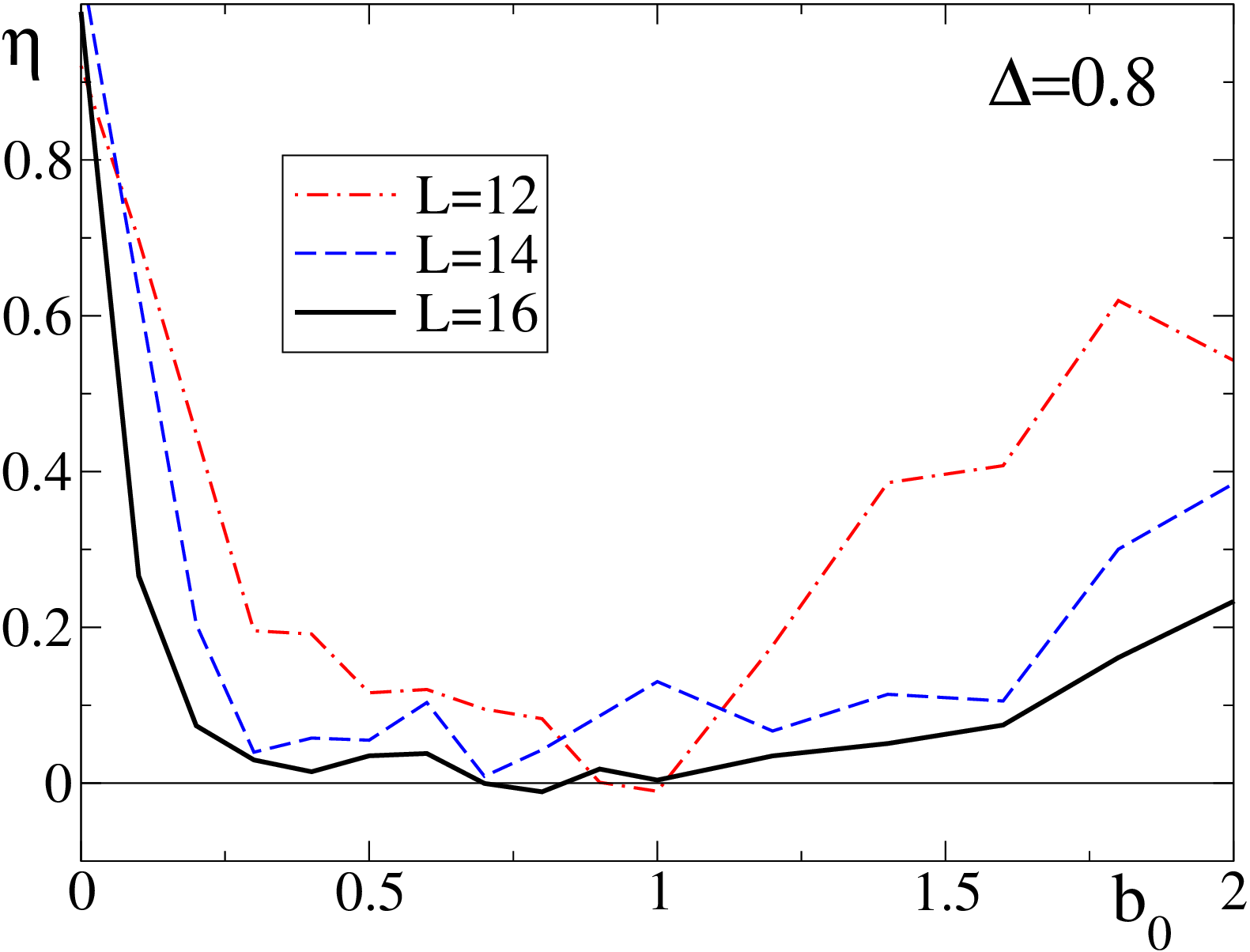}
\caption{Parameter $\eta$ for the deviation from the WD level
  distribution vs. impurity field $b_0$ for $\Delta=0.8$ and various
  $L$.}
\label{fig1}
\end{figure}

Even stronger probe of the level statistics is the correlation
$\Delta_3$ measuring the level fluctuations beyond the nearest
neighbor levels \cite{brod},
\begin{equation}
\Delta_3 = \frac{1}{2N} \mathrm{min}_{A,B} \int_{-N}^N
[{\cal N}(\tilde E)-A\tilde E-B]^2 d\tilde E , \label{delta3}
\end{equation}  
where ${\cal N}(\tilde E)$ is the integrated density of states with
$\tilde E=E/\Delta_0$ \cite{mulh}.  $\Delta_3$ should behave as
$\Delta_3 \sim N/15$ for Poisson distribution, and asymptotically as
$\Delta_3 \sim (\ln N)/\pi^2$ within the RMT \cite{brod}.  In Fig.~2
we present results for $\Delta_3(N)$ for fixed $\Delta=0.8, b_0=0.8$
as obtained for different $L=12-16$. A comparison with the result
expected from the RMT shows that $\Delta_3(N)$ approaches the latter
very accurately in an interval $N<N^*(L)$ with $N^*$ strongly
(exponentially) increasing with $L$, while the deviation into a
Poisson-like linear dependence $\Delta_3 \propto N$ appears for
$N>N^*(L)$. Such a generic crossover has been observed also in other
systems \cite{stas} and one can discuss the relevance of the related
crossover energy scale $\epsilon=N^*\Delta_0$. Fast increase of
$N^*(L)$ one can understand by noting that the impurity perturbation
being $L$-independent mixes up many-body levels
\cite{jacq} within the interval $\epsilon$ whereby separation between
many-body levels decreases as $\Delta_0 \propto \exp(-L)$. We can
estimate $\epsilon \sim b_0/(4L)$ within the XY ($\Delta=0$) model
which gives right order of magnitude for observed $N^*$ in
Fig.~2. More detailed analysis in analogy to other systems \cite{jacq}
is difficult due to the complicated nature of states at intermediate
$\Delta$.

\begin{figure}[htb]
\includegraphics[angle=0, width=.75\linewidth]{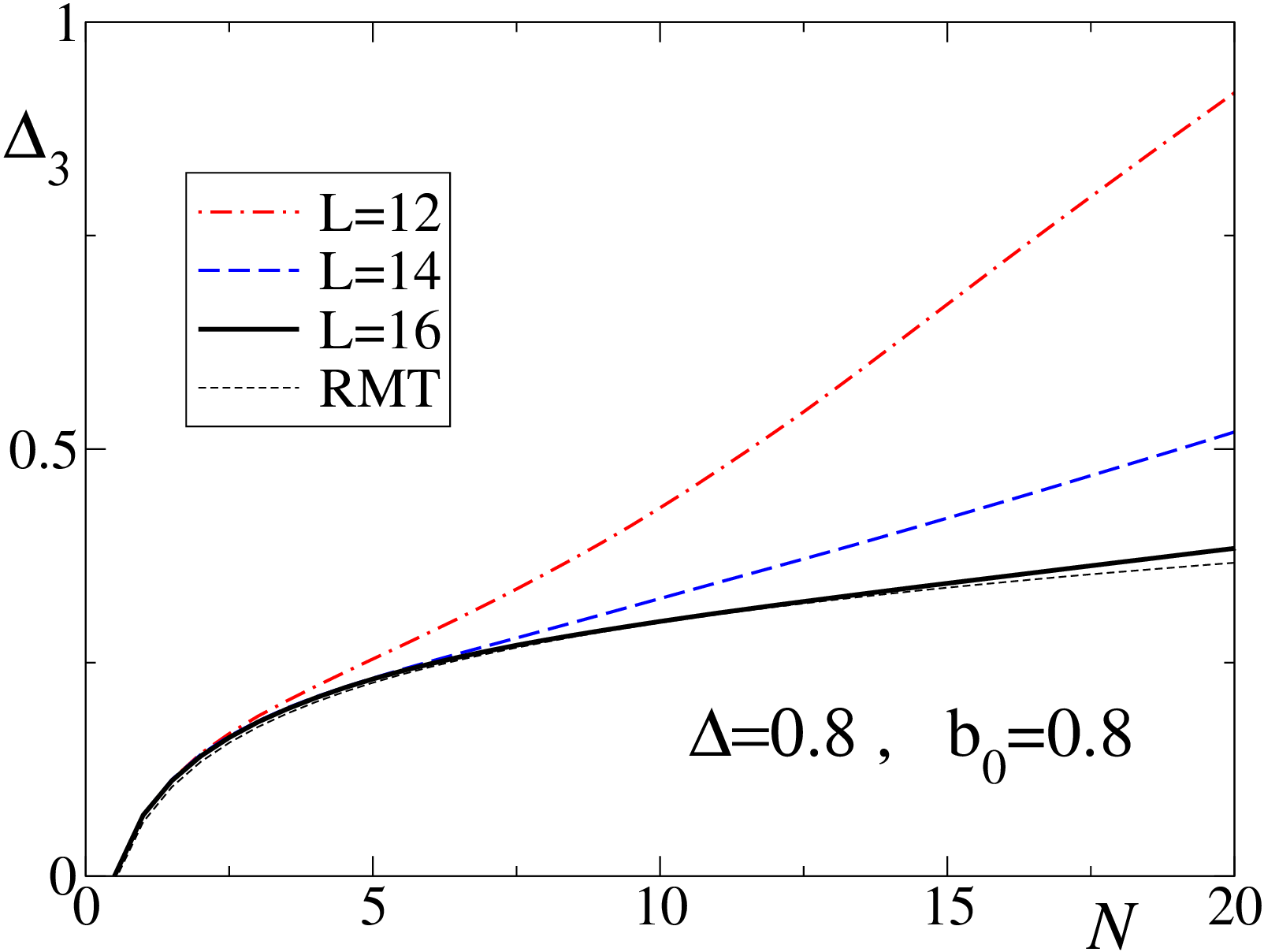}
\caption{Level-fluctuation parameter $\Delta_3(N)$ for fixed $\Delta=0.8,
  b_0=0.8$ and different system length $L$. For comparison the RMT
  result is presented (dotted line).}
\label{fig2}
\end{figure}

Closely related to the onset of the WD distribution by a single
impurity is the vanishing of the $T>0$ coherent (ballistic) transport
characteristic for integrable systems \cite{cast,znp}. The measure of
the coherent component is for the spin transport the spin
stiffness $D(T)$ (equivalent to the charge stiffness for the related
fermionic model).  It can be defined via the gauge phase $\phi$ into
the spin-flip terms in Eq.~(\ref{ahm}), as $\exp(i\phi)
S^+_{l+1}S^-_l+\exp(-i\phi) S^-_{l+1}S^+_l$. At finite $T>0$ the spin
stiffness can be expressed as
\begin{equation}
D = \frac{1}{2L} \sum_n p_n 
\frac {\partial ^2 \epsilon_n (\phi) }{\partial \phi^2}
\sim \frac{\beta}{2L} \sum_n p_n \bigl(\frac {\partial \epsilon_n (\phi) }
{\partial \phi}\bigr)^2,
\label{stiff}
\end{equation}
where $p_n=\exp(-\beta \epsilon_n)/Z$ with $Z=\sum_n\exp(-\beta
\epsilon_n)$, and the last relation becomes an equality provided that
the susceptibility for persistent current vanishes (for finite
systems at large enough $T$). On the other hand, $D$ still depends on
the value $\phi$ where derivatives in Eq.~(\ref{stiff}) are taken. For
the XY model at $\Delta=0$ corresponding via the Wigner-Jordan
transformation to tight-binding noninteracting fermions with $t=J/2$
and a potential impurity $\epsilon_0=b_0/2$ one can establish the
relation with the transmission through the barrier, as used also in
connection with the evaluation of the 1D conductance
\cite{reje} at $T=0$.  For general $T>0$ one gets in the case of NI
fermions and $L \to \infty$,
\begin{equation}
D= \frac{\beta}{2L} \sum_k f_k (1-f_k) (v_k)^2 g_k. \label{dni}
\end{equation}
where $v_k =2t \sin k$, $f_k$ is the Fermi function with $\mu=0$ for 
half-filling ($S^z_{tot}=0$) and 
\begin{equation}
g_k= \frac{|t_k|^2\sin^2(L\phi)}{1-|t_k|^2\cos^2(L\phi)},
\qquad |t_k|^2=\frac{4t^2 \sin^2 k}{4t^2 \sin^2 k+\epsilon_0^2}.
\label{gk}
\end{equation}
Numerically we recover the behavior $D(\phi)$ as follows from
Eqs.~(\ref{dni},\ref{gk}) for arbitrary $b_0$ as far as $\Delta \to
0$. For $\Delta>0$ the dependence on $\phi$ remains qualitatively
similar, although irregular due to strong dependence on $L$.  In the
following we calculate $D(L)$ for fixed $\phi=\pi/(2L)$. Results for
$\Delta>0$ are nontrivial for any $T$. Since results of full ED are
best at high $T$, we restrict ourselves here to the limit $\beta \to
0$. It has been shown for the pure model that $D/\beta$ remains finite
and nontrivial in the thermodynamic limit $L \to \infty$ due to
integrability of the model \cite{znp}.

In Fig.~3 we show results for $D/\beta$ vs. $1/L$ for chosen
$\Delta=0.8$ and for four cases $b_0=0, 0.5, 1, 2$. It is evident from
Fig.~3 that $b_0>0$ cases are qualitatively different from the $b_0=0$
where $D$ scales linearly in $1/L$ towards a finite $D/\beta
\sim 0.035$. On the other hand, $b_0>0$ induces an exponential-like
decay of $D \to 0$, at least for large enough $L>L^*$ and not too
weak $b_0$. This is closely related to the onset of the WD
distribution and the effective breaking of the integrability.

\begin{figure}[htb]
\includegraphics[angle=0, width=.75\linewidth]{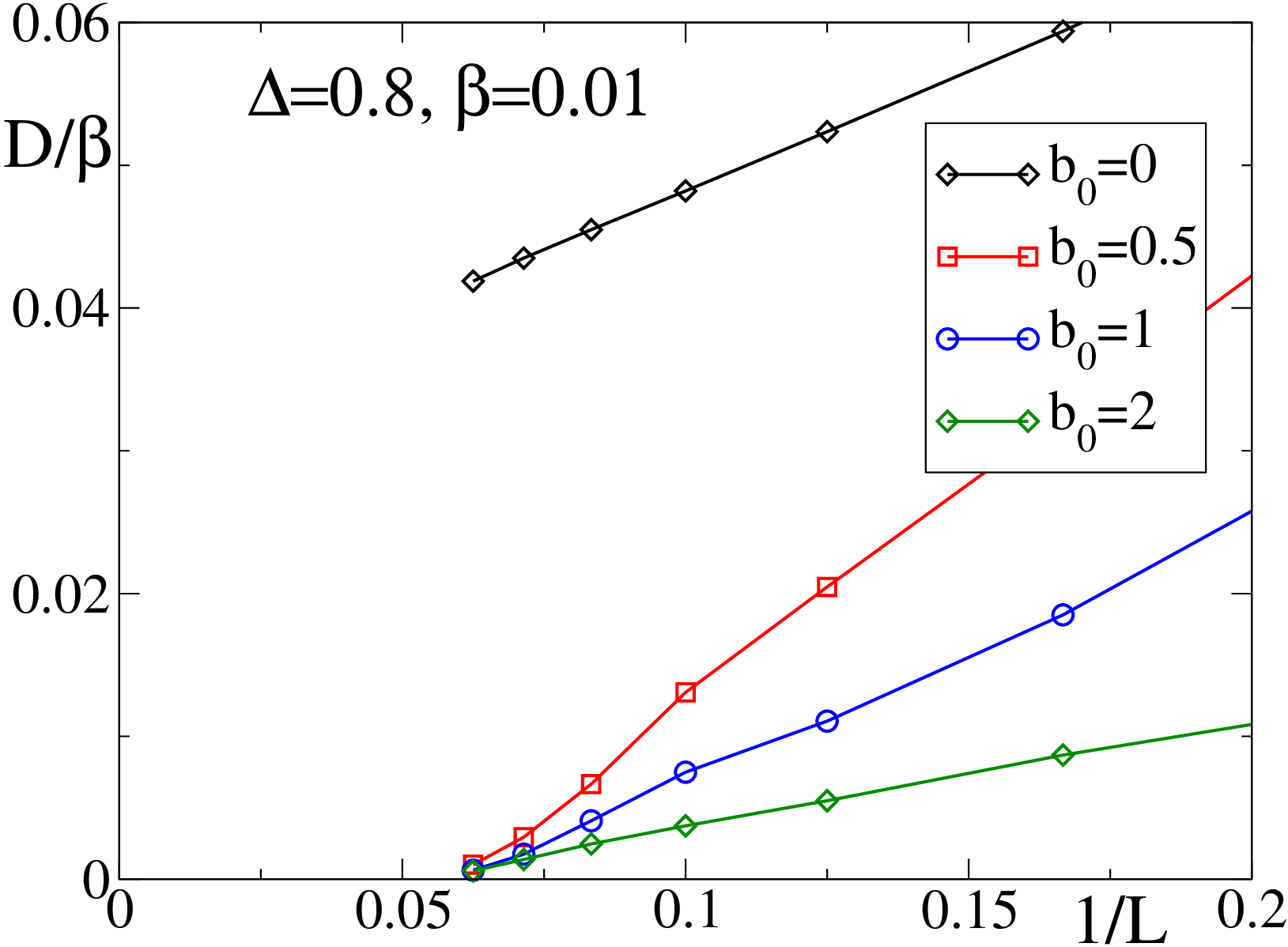}
\caption{High-$T$ spin-stiffness $D/\beta$ vs. $1/L$ for fixed
  $\Delta=0.8$ and different $b_0$.}
\label{fig3}
\end{figure}

Rapid (exponential) vanishing of $D(L \to \infty)$ at $T>0$ is the
indication that the transport is not ballistic and becomes incoherent
(resistive) beyond the characteristic $L^*$.  In order to test this
directly we evaluate dynamical spin conductivity $\sigma(\omega)$ as
well as the related thermal conductivity $\kappa(\omega)$, defined as
\begin{equation}
\sigma(\omega)=\frac{i(\chi_{jj}^0-\chi_{jj}(\omega))}{\omega^+L}, 
\quad \kappa(\omega)=\frac{i\beta(\chi_{j_Ej_E}^0-
\chi_{j_Ej_E}(\omega))}{\omega^+ L}, \label{sig}
\end{equation}
where $j$ and $j_E$ are spin and energy current, respectively, with
corresponding susceptibilities 
\begin{equation}
\chi_{jj}(\omega)=i\int_0^\infty dt e^{i\omega^+ t} \langle[j(t),j]
\rangle,   \label{chi}
\end{equation}
and analogous definition of $\chi_{j_Ej_E}(\omega)$. Note that for
'normal' transport one expects $\chi_{jj}^0=\chi_{jj}(\omega \to 0)$.
In a nondissipative case, however, $\chi_{jj}^0-\chi_{jj}(\omega \to
0))=2 L D>0$. For further discussion it is convenient to introduce and
analyze also corresponding memory functions $M(\omega)$
and $N(\omega)$, defined respectively as \cite{gotz}
\begin{equation}
\sigma(\omega)=\frac{i}{L} \frac{\chi_{jj}^0}{\omega+ M(\omega)},
\qquad \kappa(\omega)=\frac{i\beta}{L} \frac{\chi_{j_Ej_E}^0}{\omega+
  N(\omega)}. \label{mem}
\end{equation} 

The advantage of studying $\kappa(\omega)$ is that $j_E$ is a
conserved quantity in the pure AHM \cite{znp}, hence $N_0(\omega)=0$
and consequently $N(\omega) \ne 0$ appears only due to $b_0
\ne 0$. On the other hand, $j$ is not conserved and
$M(\omega)=M_0(\omega)$ is nontrivial even in the absence of
impurities. Nevertheless, $M''_0(\omega=0)=0$ at any $T$ as required to
obtain $D(T)>0$. In the following we evaluate $\sigma(\omega)$ and
$\kappa(\omega)$ at $T>0$ using the microcanonical Lanczos method
(MCLM) \cite{long} to calculate the dynamical susceptibilities,
Eqs.~(\ref{chi},\ref{sig}), in systems with $L=16-24$. Typically, $N_L
\sim 2000$ Lanczos steps are used to obtain spectra with high $\omega$
resolution, so that the additional broadening is only $\delta = 0.01$.
In the following we present merely results at $\beta \to 0$, which do
not qualitatively change with $T$ down to $T \sim J$. At least for
larger $L$ smooth $\sigma(\omega)$ and $\kappa(\omega)$ then allow the
evaluation of $M(\omega)$ and $N(\omega)$ via Eq.~(\ref{mem}).

\begin{figure}[htb]
\includegraphics[angle=0, width=.85\linewidth]{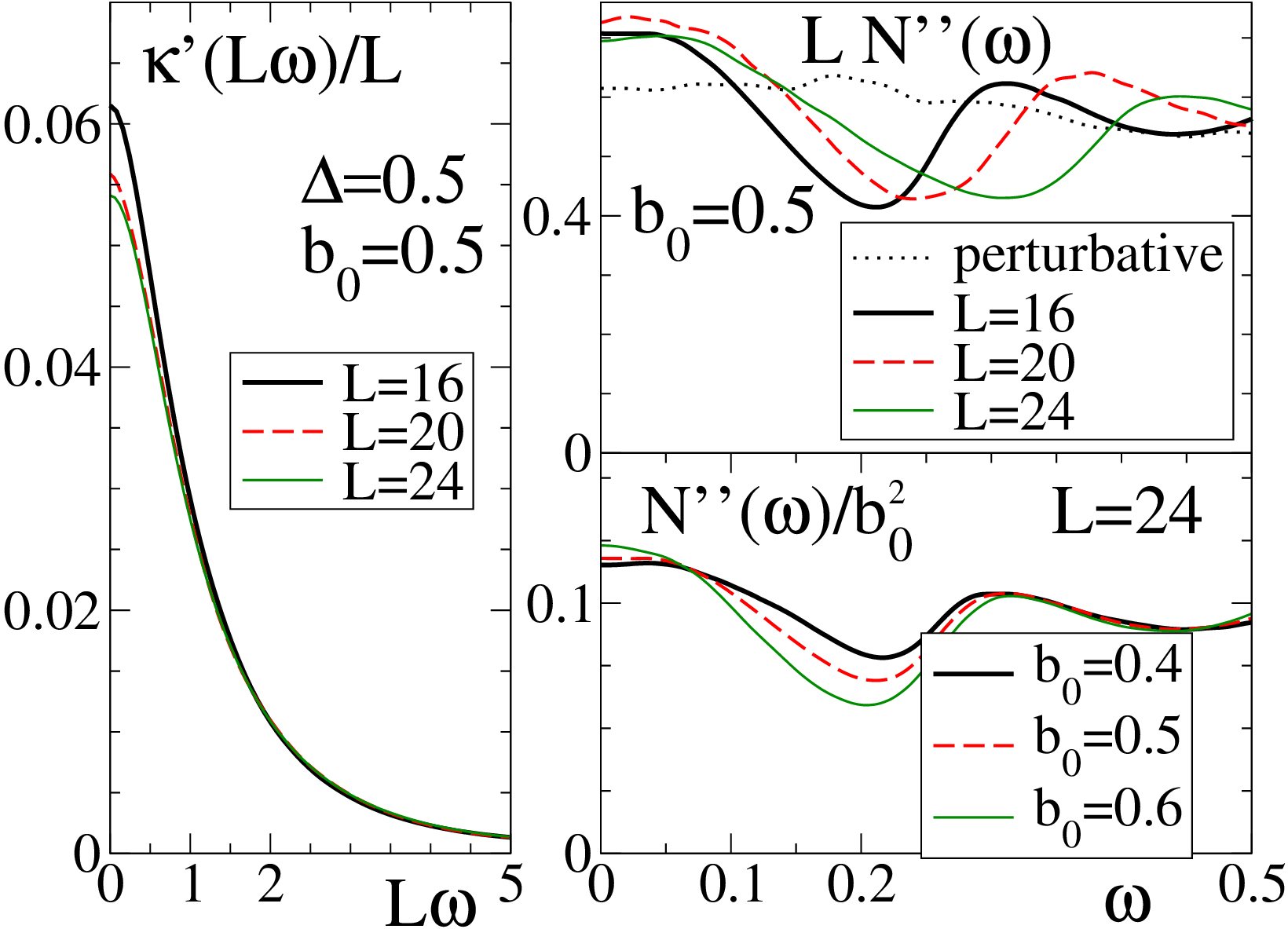}
\caption{a) High-$T$ results for scaled thermal conductivity
  $\kappa'/L$ vs. $L \omega$ for $\Delta=0.5$ and $b_0=0.5$ as
  calculated for different $L=16-24$. b) Extracted scaled memory
  function $L N''(\omega)$. Perturbation-theory result is also plotted
  (dashed curve). c) $N''(\omega)/b_0^2$ for $\Delta=0.5$ and
  different impurity fields $b_0=0.3-0.5$, obtained at fixed $L=24$.}
\label{fig4}
\end{figure}

Let us start with the analysis of (real part) $\kappa'(\omega)$, which
reveals a Lorentzian (Drude) form for $\Delta=0.5, b_0=0.5$ as shown
in Fig.~4a. Moreover, results show universal size scaling as
$\kappa'(L\omega)/L$. Hence, corresponding $N''(\omega)$ also scales
as $1/L$, so that the size-independent quantity is $\tilde
N(\omega)=LN(\omega)$ being quite structureless for $\omega<1$ as
shown in Fig.~4b. It is also evident from Fig.~4c that $N''(\omega)
\propto b_0^2$ at least for weaker $b_0<0.5$.  On the other hand, at
larger $b_0>0.5$ $N''(\omega)$ obtains a characteristic peak at low
$\omega \to 0$ which strongly reduces the d.c. value
$\kappa(\omega=0)$. I.e., on entering the regime $b_0>1$ the impurity
starts to cut the ring for the d.c. transport.

The regular behavior of $N''(\omega)$ for weaker $b_0<1$ gives support
to the attempt to evaluate the memory function within the perturbation
approach \cite{gotz} using the force-force correlations,
\begin{equation}
N_p(\omega) = \frac{1}{\omega \chi^0_{j_Ej_E}}
(\chi_{ff}(\omega)-\chi_{ff}^0), \qquad
f=i[H,j_E], \label{gotz}
\end{equation}
where $\chi_{ff}(\omega)$ is the force-force dynamical
susceptibility. Results for $N_p''(\omega)/L$ evaluated using
eigen-states obtained by the ED of the system $L=14$ are for comparison
also presented in Fig.~4b. The correspondence is quite satisfactory.

\begin{figure}[htb]
\includegraphics[angle=0, width=.75\linewidth]{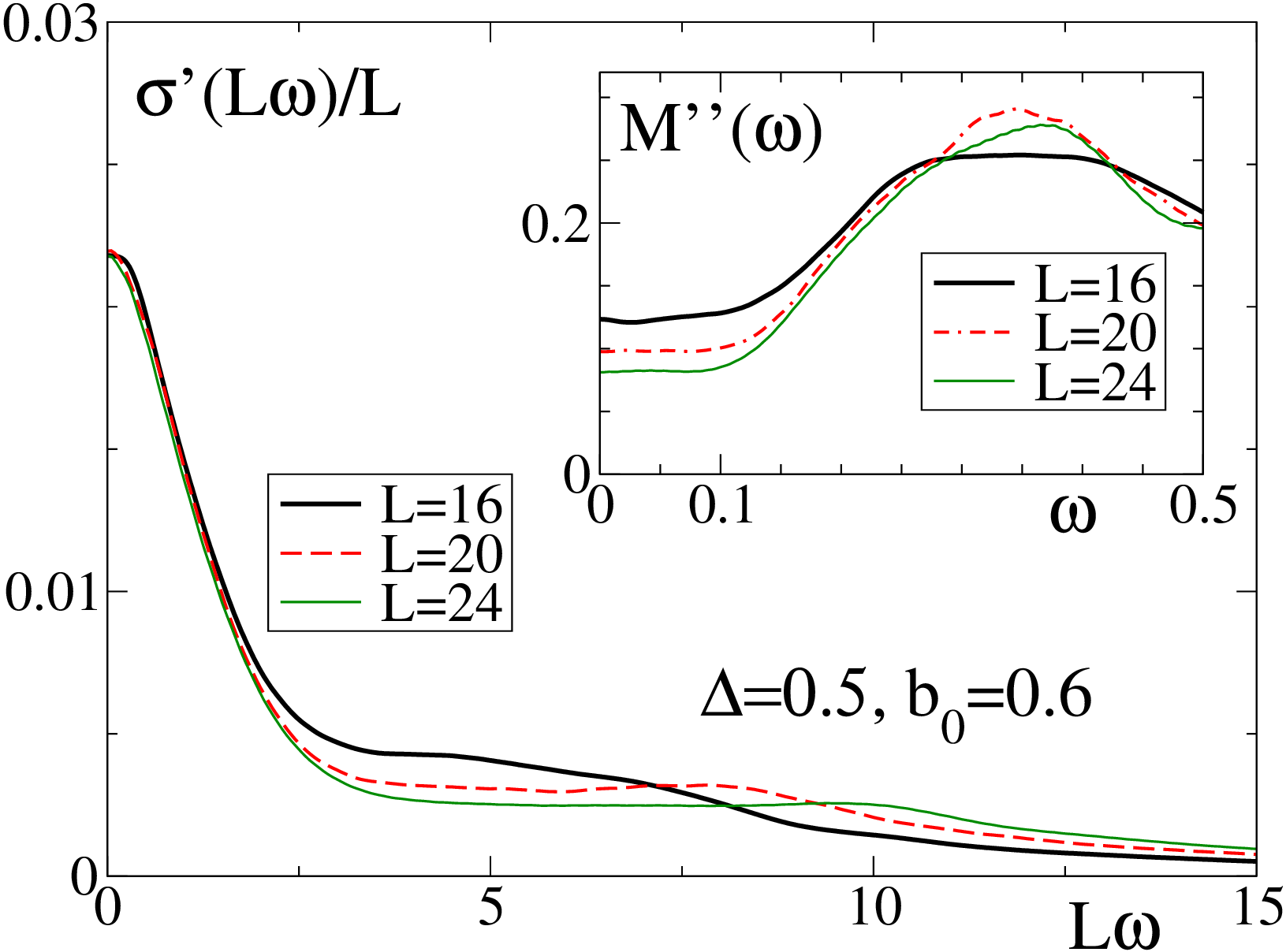}
\caption{a) Scaled spin conductivity $\sigma'/L$ vs. $L\omega$ for
  $\Delta=0.5, b_0=0.6$ for different $L=16-24$, and b) corresponding
  $M''(\omega)$. }
\label{fig5}
\end{figure}

Finally, we present in Fig.~5 also analogous scaled $\sigma'/L$
vs. $L\omega$ and corresponding $M''(\omega)$ for fixed
$\Delta=0.5,b_0=0.6$ and different $L=16-24$, as obtained via the ED
($L=16$) and MCLM ($L=20,24$) at $\beta \to 0$. Since $M_0(\omega)$ is
nontrivial even for $b_0=0$ one can discuss possible decomposition
$M(\omega)=M_0(\omega)+\tilde M(\omega)/L$. Results confirm that at
low $\omega<0.2$ $\sigma(\omega)$ reveals a Lorentzian with $M''(\omega
\to 0)$ scaling as $1/L$, the only contribution in this regime coming
from the impurity. Moreover, we notice that for fixed $\Delta<1$ and
$b_0$ $M''(0)$ and $N''(0)$ are quite similar in value, e.g., by
comparing Fig.~4c and Fig.~5. This indicates that we are at $\Delta<1$
close to the validity of the Wiedemann-Franz law requiring an unique
transport relaxation rate.

It should be pointed out that obtained incoherent transport is
characterized with the relaxation times $\tau$ and d.c. conductivities
scaling linearly with $L$, as expected for 1D systems with a single
perturbed region. Hence, the length independent quantities are
$\sigma(0)/L=\chi^0_{jj}/(LM''(\omega=0))= \chi^0_{jj}/\tilde
M''(\omega=0)$ and the corresponding thermal one $\kappa(0)/L=\beta
\chi^0_{j_Ej_E}/\tilde N''(\omega=0)$. 

In conclusion, we have shown that the transport in the considered
anisotropic Heisenberg model on the ring with a single static impurity
is quite unique.  Since both the spin (for $\Delta<1$) and thermal
conductivity at any $T>0$ are dissipationless in the pure system, one
can study directly the nontrivial effect of a single impurity on the
level statistics and transport in the many-body quantum system. We
have shown that single static impurity induces an incoherent transport
with a well defined current relaxation time which scales as $\tau
\propto L$. This should be contrasted with the case of noninteracting
fermions in Eq.~(\ref{dni}) where a single impurity only reduces the
stiffness $D$ but does not lead to the current relaxation within the
ring at any $T$.  The fundamental difference seems to come from the
Umklapp processes which are revived by the impurity and lead to the
decoherence between successive scattering events on the impurity. In
this sense it is also plausible that for a finite but low
concentration $c_i$ of static impurities in a chain (as relevant for
experiments \cite{hess}) we expect that our results can be simply
generalized as $1/\tau \propto 1/L \to c_i$, as evident also from the
lowest-order perturbation theory, Eq.~(\ref{gotz}).

Although we studied here the AHM model, results and conclusions could
be plausibly generalized to 1D integrable chain systems with periodic
boundary conditions and a localized perturbed region. In this paper we
presented only results in the high-$T$ regime, still the phenomenon is
expected to persist as far as the Umklapp processes are effective,
i.e., for $T$ above some characteristic Umklapp temperature $T_U$. We
can speculate that the LL phenomenology \cite{kane,furu} can become
effective only for $T \ll T_U$, whereby there is another Kane-Fisher
scale $T^*$ which divides the regimes of cut chain for $T<T^*$ and
renormalized coherent transmission for $T>T^*$. Further studies are
needed to establish these scales for relevant impurities and
models. In any case, such phenomena are expected to be relevant in
connection with recent experiments on thermal conductivity in spin
chains where dilute impurities are introduced \cite{hess}.

This work was supported by the FP6-032980-2 NOVMAG and COST P-16 ECOM
project and by the Slovenian Agency grant No. P1-0044.

\end{document}